\documentclass [twocolumn,epsfig,10pt,floatfix,amsmath,amssym]{revtex4}
\usepackage[dvips]{epsfig}

\usepackage{float}

\begin{document}

\title{Interfaces in Diblocks: A Study of Miktoarm Star Copolymers}

\author{Gregory M. Grason}
\author{Randall D. Kamien}

\affiliation{Department of Physics and Astronomy, University of Pennsylvania,
Philadelphia, PA 19104-6396, USA}
\begin{abstract}
We study AB$_n$ miktoarm star block copolymers in the strong segregation limit, focussing on the role that the AB interface plays in determining the phase behavior.  We develop an extension of the kinked-path approach which allows
us to explore the energetic dependence on interfacial shape.  We consider a one-parameter family of interfaces to study the columnar to lamellar transition in asymmetric stars. We compare with recent experimental results.  We discuss the stability of the A15 lattice of sphere-like micelles in the context of interfacial energy minimization.  We corroborate our theory by implementing a numerically exact self-consistent field theory to probe the phase diagram and the shape of the AB interface.  
\end{abstract}
\date{\today}

\maketitle

\section{Introduction}

Not only are block copolymers promising materials for nano-patterned structures \cite{Chaikin, Russell}, drug delivery \cite{Discher}, and photonic applications \cite{thomas_prb_02}, but they are also the 
ideal system for studying the
influence of molecule architecture on macromolecular self-assembly \cite{bates_arpc}.  Because of the ongoing interest in novel macromolecular organization, theoretical predictions based on heuristic characterization of molecular architecture offer crucial guidance to synthetic, experimental, and theoretical studies.  Though the standard diblock copolymer phase diagram \cite{Leibler} was explained nearly a quarter of a century ago, the prediction and control of phase boundaries is fraught with subtle physical effects:
weak segregation theory provides an understanding of the order-disorder transition \cite{fred_helf_jcp_87}, strong segregation theory (SST) predicts most of the ordered morphologies \cite{semenov}, and numerically exact, self-consistent field theory (SCFT) \cite{mat_sch_prl} can resolve the small energetic differences between
a variety of competing complex phases.  

In previous work, we argued that in diblock systems, that as the volume fraction of the inner block grows, AB interfaces are deformed into the shape of the Voronoi polyhedra of micelle lattice, and therefore, the free-energy of micelle phases can be computed simply by studying properties of these polyhedra.  In particular, we predicted that as volume fraction of inner micelle domain grows the A15 lattice of spheres should minimize the free energy as long as the hexagonal columnar phase (Hex) did
not intervene \cite{gra_didon_kam_03}.  We corroborated this prediction by implementing a spectral SCFT \cite{mat_sch_prl} for branched diblock copolymers: in this paper we probe the regime of validity of our analytic analysis through both
strong segregation theory and SCFT.   Though there is extremely small variation in the energy between different interfacial geometries, so too is the variation in energy between different stable phases.  Thus, we compare these two approaches not only by the phase diagram but also through the details of the ordering in the mesophases.  Since our original {\sl ansatz} hinged on the (minimal) area of the interface between the incompatible
blocks, we will focus strongly on the shape and structure of this interface. We will explore in detail the relationship between molecule architecture and the polyhedral distortion of the AB interface induced by the lattice packing of micelles to study hexagonal columnar
phases.  Our results motivate the search for a stable A15 phase which we find in SCFT. 

\begin{figure}
\center
\epsfig{file=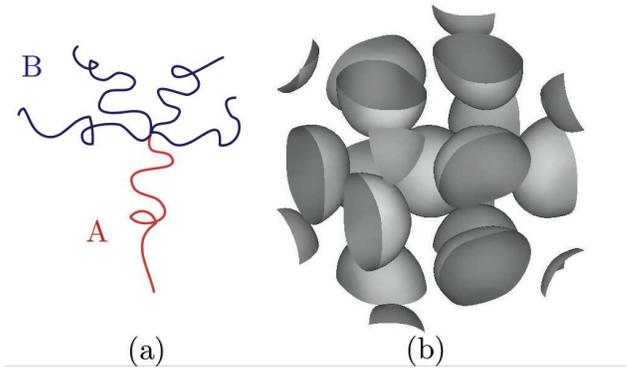,width=3.25in} \center \caption{The miktoarm star architecture for an AB$_4$ diblock copolymer is shown in (a).  Shown in the $Pm\bar{3}n$ unit cell, (b), are the AB interfaces for the A15 phase, extracted from SCFT results for $n=5$ at $\chi N =40$ and $f=0.349$ (along the Hex-A15 phase boundary).   The A15 unit cell contains 8 micelles:  2 at the center and corners (as with BCC); and 6 along the faces of the unit cube. }
\label{fig: figure1}
\end{figure}

In order to render the sphere-like phases stable in comparison to the Hex phase, we are obliged to consider asymmetric diblocks;
while symmetric, linear diblock copolymers with an A and a B block
have equivalent ``inside-out'' 
bulk morphologies when the A volume fraction $f$ is replaced with the B volume
fraction $(1-f)$, copolymers with
branched or otherwise asymmetric architectures have no such
symmetry, and therefore, tend to favor morphologies with one of
the two components on the ``outside" a of curved AB interface
(i.e. on the outside of micelles).  Indeed, our previous SCFT analysis of branched diblocks is consistent
with these findings. Because of the challenge of accounting for all the competing effects, in Section II we implement a full SCFT for diblocks with the AB$_n$ architecture to explore the mean field behavior of mitkoarm melts.
In Section III we develop a strong-segregation theory approach for the hexagonal columnar  phase which allows us to parameterize a large class of configurations and to explicitly assess the accuracy of the unit-cell approximation (UCA), which assumes the lattice Voronoi cell to be perfectly cylindrical (or spherical for three-dimensional lattices). Our calculation builds on the ``kinked-path" calculation of Milner and Olmsted \cite{olm_mil_macro_98}, and allows us to explore the influence of the hexagonal micelle lattice on the cylindrical morphology.  We find that the shape of the Voronoi cell of the lattice strongly influences the shape of the AB interface.   In Section IV we compare the predictions of the full SCFT calculation to the SST calculation in order to assess the accuracy of the latter.   In addition, we demonstrate how the SST results of Section III can be used to compute an accurate phase boundary for transitions between lamellar (Lam) to Hex configurations.  We briefly discuss the inverse phases (where the B blocks are on the inside) in Section V. Finally, we conclude in Section VI.

\section{Self-Consistent Field Theory of AB$_n$ Miktoarm Star Copolymers}

Approximate self-consistent field theory calculations have explored the mean field phase behavior of linear diblocks with asymmetric monomer sizes \cite{vavasour_macro_93} which were confirmed through numerically exact SCFT \cite{mat_sch_macro_94_1,mat_bates_jpsci}.  Milner
developed SST, applicable in the
$\chi N \rightarrow \infty$ limit ($\chi$ is the Flory-Huggins parameter for A and B monomers and $N$ is the degree of polymerization of the copolymers), for melts of A$_n$B$_m$
miktoarm star copolymers which also incorporates asymmetric monomer sizes \cite{milner_macro_94,
olm_mil_macro_98}.  Generally, the results of all of these
calculations show that equilibrium morphologies which have blocks
with stronger effective spring constants (i.e. more arms or
smaller statistical segment lengths) on the outside of curved
interfaces are favored over a much larger region of the phase
space than in the symmetric diblock case.  
The details of the calculation implemented here will be reported elsewhere as a specific case of more general SCFT calculation for multiply-branched diblocks \cite{grason_kamien_tobe}.  The method is an extension of Matsen and Schick's spectral SCFT calculation for melts of linear \cite{mat_sch_prl, mat_sch_macro_94_1} and starblock copolymers \cite{mat_sch_macro_94_2}.
Given the space group of a copolymer configuration, the mean field free-energy can be computed to arbitrary accuracy.  The results of these SCFT calculations are accurate to the extent that mean field theory is correct and composition fluctuations can be ignored.  The contributions of these fluctuations tends to zero in the $N \rightarrow \infty$ limit \cite{fred_helf_jcp_87}, and therefore we can expect to capture the equilibrium results observed in the PS-PI miktoarm star experiments, for which $N \sim 1,000$ \cite{beyer_jpsci_99, pochan_macro_96, tselikas_jchemphys_96, beyer_macro_99, yang_macro_01}.

We consider a melt of fixed volume and number of copolymers, with each molecule composed of $N$ total monomers.  The volume fraction of the A-type monomer is $f$.   In general, we could accomodate monomer asymmetry by allowing for two different statistical segment lengths, $a_A$ and $a_B$, for the A and B species, respectively.   In this case, each statistical segment length can be scaled appropriately so that the physical ``packing length", $\ell_{A,B} = \rho_{A,B}^{-1}/a_{A,B}^2$, is fixed to the proper value for each of the two chemical species \cite{milner_macro_94}.  Thus, without loss of generality we define a common segment density for the two monomer types, $\rho_0$.  Moreover, for our SCFT calculations in this section we will restrict ourselves to the case $a_A=a_B$.  The asymmetry of the copolymers we study arises entirely from their architecture.  Each copolymer is composed of one block of pure A monomer joined to $n$ blocks of pure B monomer at a common junction point (see Figure \ref{fig: figure1} (a)).

\subsection{Low to Intermediate Segregation}

\begin{figure}
\center
\epsfig{file=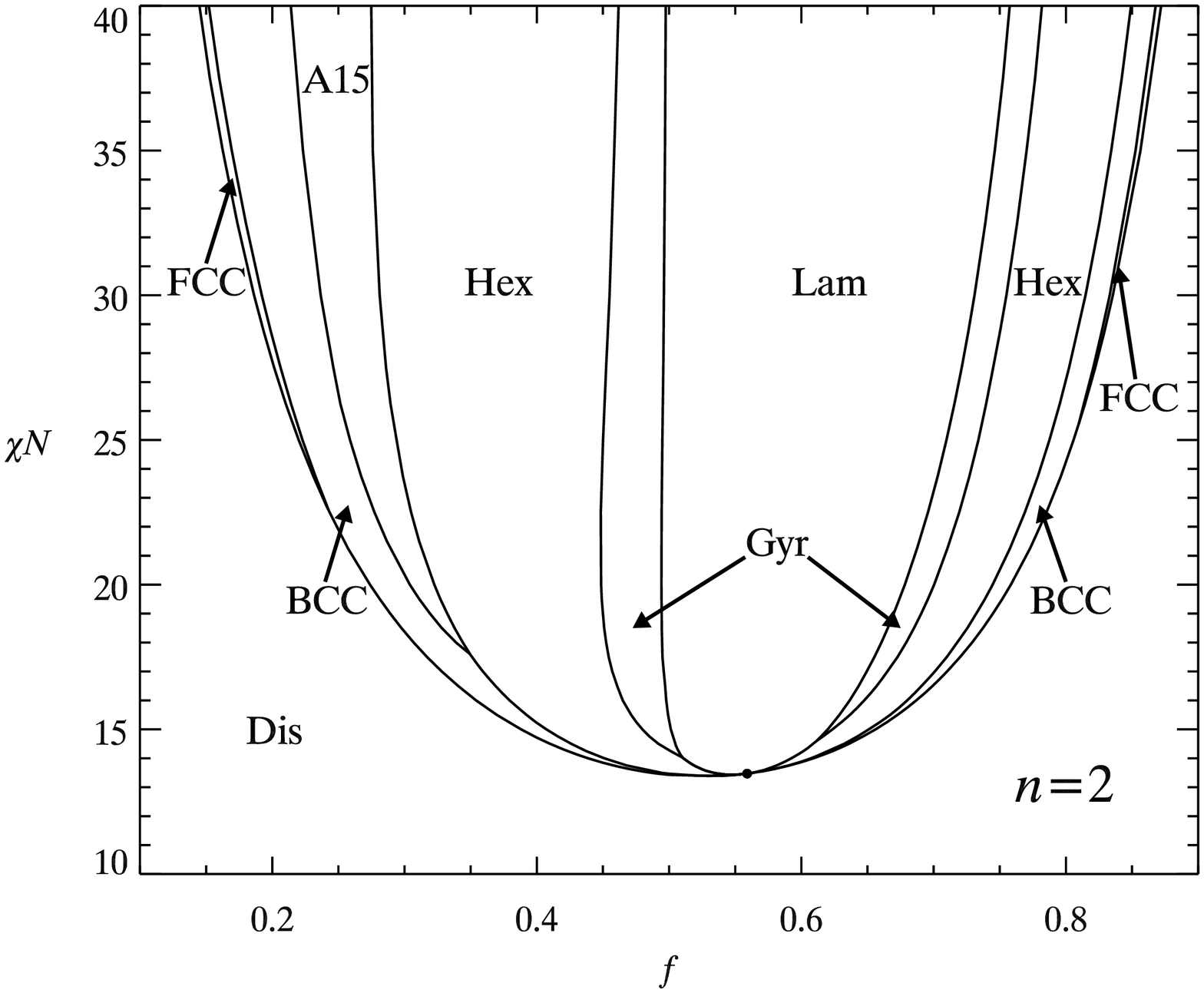,width=3.5in}\center
\epsfig{file=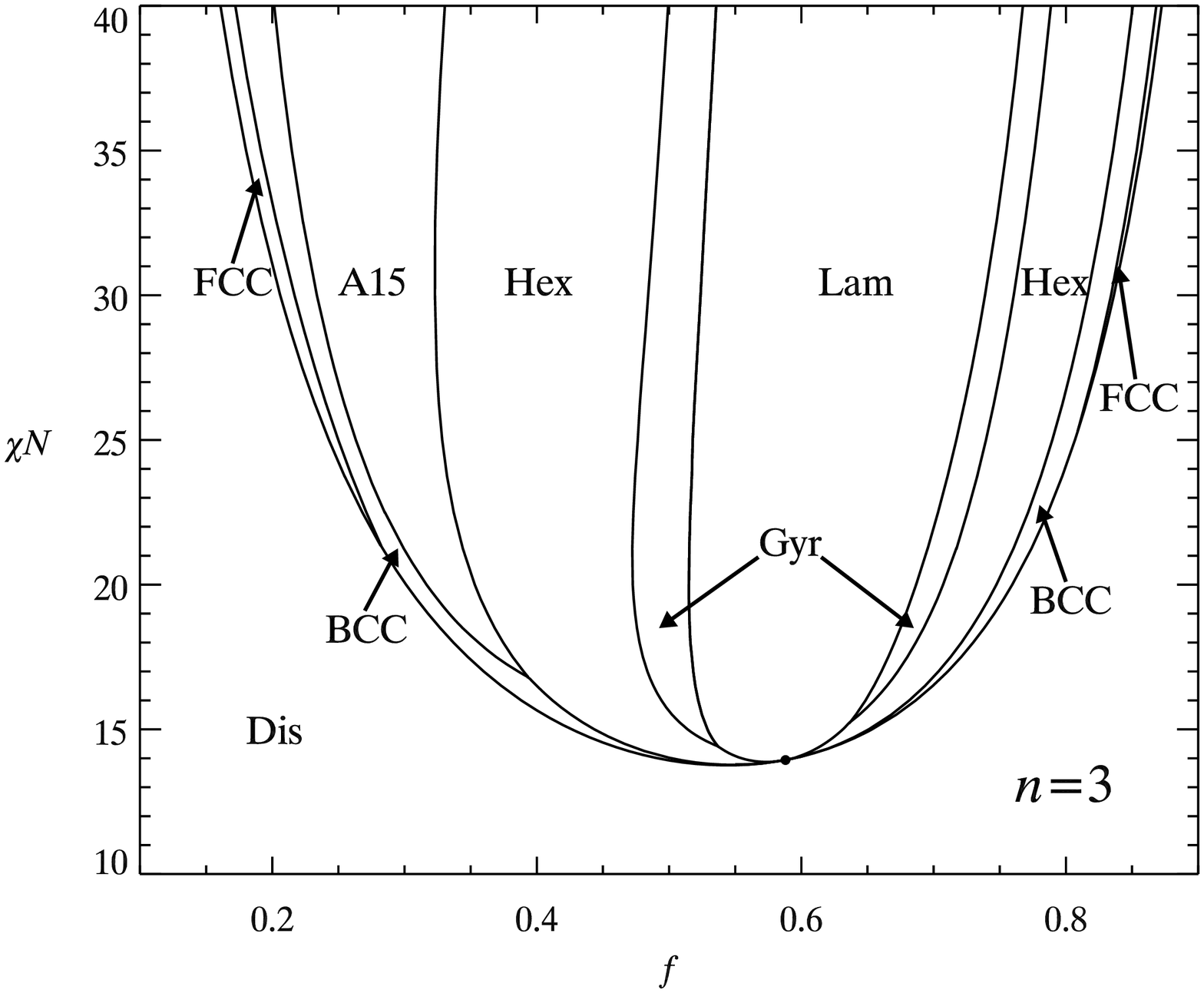,width=3.5in}\center
\caption{Phase diagrams for $n=2$ and $n=3$. Dis labels regions where the melt is disordered. Stable regions of ordered phases are labeled:  (Lam) lamellar;  (Gyr) gyroid, $Ia\bar{3}d$ symmetry; (Hex) hexagonal-columnar, $p6mm$ symmetry; (A15) sphere phase, $Pm\bar{3}n$ symmetry; (BCC) body-center cubic lattice of spheres, $Im\bar{3}m$ symmetry; and (FCC) face-centered cubic lattice of spheres, $Fm\bar{3}m$ symmetry \cite{FCC_note}.  The circle marks the mean field critical point through which the system can transition from the disordered state to the Lam phase via a continuous, second-order phase transition.  All other phase transitions are first-order.} \label{fig: figure2}
\end{figure}

\begin{figure}\center
\epsfig{file=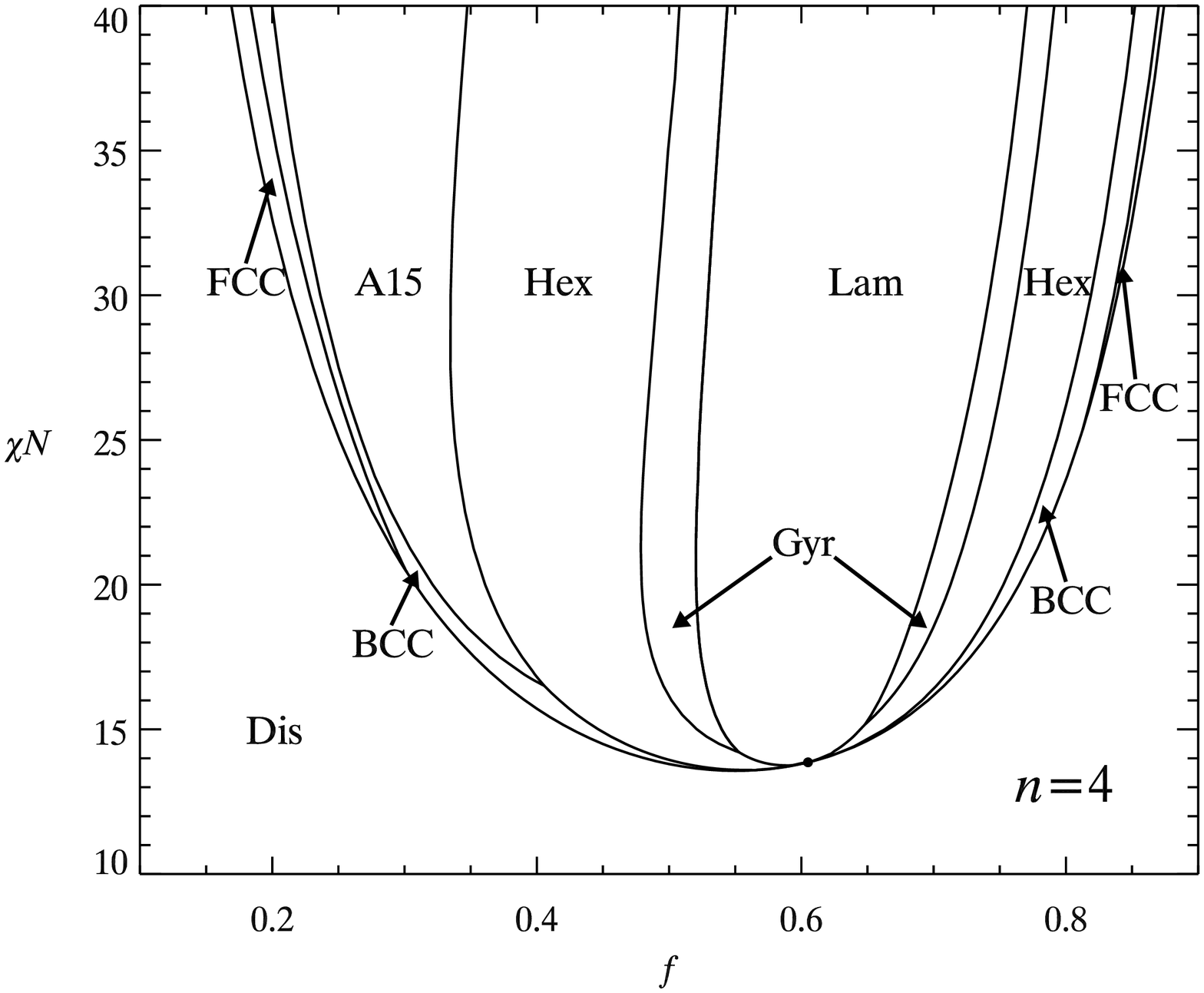,width=3.5in}\center
\epsfig{file=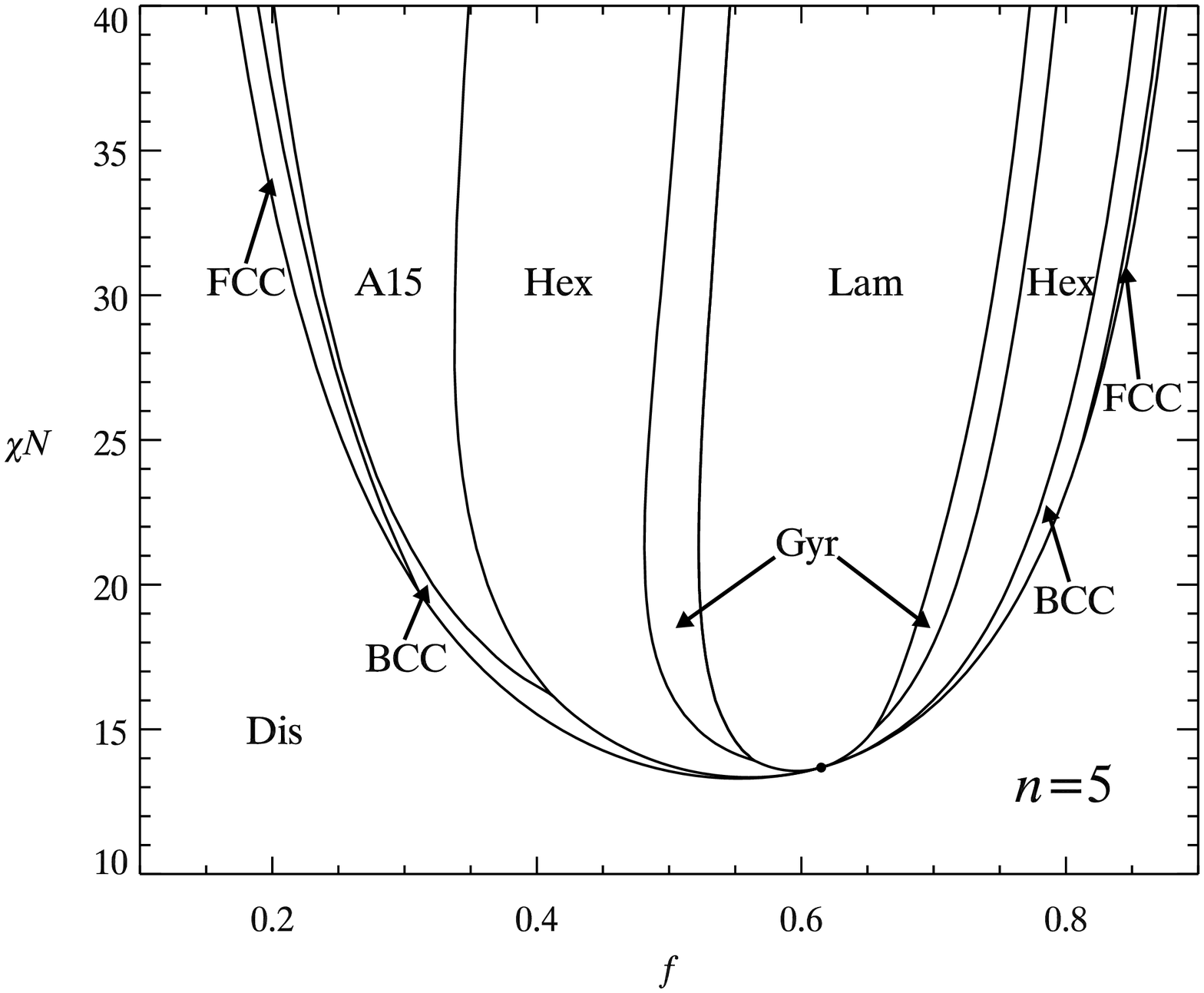,width=3.5in}\center
\caption{Phase diagrams for $n=4$ and $n=5$.  Stable phases and critical points are labeled as in Figure \ref{fig: figure2}. } \label{fig: figure3}
\end{figure}

We computed the full phase behavior for $\chi N \leq 40$ for $n \leq 5$.  To achieve a numerical accuracy of $0.005\%$ for our free-energy calculations we employ up to 712 basis functions.  This allows a precision in the phase boundary calculations which is better that $\pm 0.001$ for $f$ and $\pm 0.01$ for $\chi N$.  The computed SCFT phase diagrams are shown in Figures \ref{fig: figure2} and \ref{fig: figure3}.

The first notable feature of these phase diagrams is that they are not symmetric about $f=0.5$, as is the case for symmetric diblocks -- phase boundaries are shifted to the right.  This indicates that phases with the $n$ B blocks on the outside of curved interfaces are favored.  By adopting an interface with non-zero mean-curvature, the configuration relaxes the blocks on the outside of the AB interface at the expense of an increase in stretching of the inner blocks \cite{matsen_jphys_02}.  Due to the additional asymmetry introduced by the molecular architecture, the stability of phases with the B blocks on the outside of highly curved interfaces is enhanced.  Moreover, this effect is generally amplified by further increasing the number of B blocks in the molecule.  If we look, for instance, at the boundary between flat and curved interfaces which separates the gyroid (Gyr) from the lamellar (Lam) phase, we see that for  $\chi N=25$ the
transition occurs at $f = 0.358$, $0.495$, $0.518$, $0.523$ and $0.525$ for $n=1$, 2, 3, 4, and $5$, respectively.  Likewise, the 
``inverted" morphologies, with B blocks on the inside of curved interfaces, are suppressed because curving the interface inward towards the B domain introduces excess stretching in these blocks.  This pushes the phase boundaries to greater $f$ for the inverse structures as well.

We also note the appearance of a stable A15 phase of spherical micelles (see Figure \ref{fig: figure1} (b)).  The A15 phase is also observed in experiments and simulations of another soft molecular system, namely neat mixtures of low molecular weight dendrons \cite{percec_jacs, percec_nature, ziherl_jpc, goddard_jacs}.  The stability of this phase has been attributed to the fact that the area of the A15 Voronoi cell is the minimal among the Voronoi cells of three dimensional periodic structures (for a given number density) \cite{ziherl_kam_prl, weaire_phelan_94}.  We have argued that this minimal area makes the A15 lattice stable in the phase diagram of a 3-generation, multiply-branched copolymer \cite{gra_didon_kam_03}:  in the (extreme) limit that the AB interface adopts the flat shape of the Voronoi cell of the micelle lattice, the A15 phase is stable over other sphere phases, such as BCC and FCC.  Thus, the appearance of the A15 phase in these phase diagrams is evidence of a highly distorted AB interface when B blocks compose the outer domains of micelles.  Due to melt incompressibilty, copolymer configurations must fill the lattice Voronoi cell so that the monomer density is uniform \cite{thomas_macro_84}.  The tension of the outer block chains stretching towards the corners of the cell pulls on the interface, distorting it towards the polyhedral shape of the Voronoi cell.  Of course, the interfacial tension of the AB interface and the tension of the inner blocks will frustrate this distortion, and the interface will actually distort into a rounded polyhedra of the same symmetry as the Voronoi cell. In the next sections we will return to this point and will show, nonetheless, that the approximation of flat-faced AB interfaces
improves as the number of outer arms grows.

Indeed, from Figures \ref{fig: figure2} and \ref{fig: figure3} it is clear that increasing the asymmetry increases the stability of the A15 phase; the window of A15 stability increases further in both the larger $f$ and smaller $f$ directions.  Since increasing $n$, the number of B blocks per molecule, stabilizes sphere-like phases over cylindrical phases for larger $f$ the A15-Hex boundary should move to
larger $f$.  Moreover, as $n$ grows the effective spring constant in the outer blocks grows as $n^2$: for fixed number of B monomers $(1-f)N$, there are $n$ entropic springs with spring constant $[a^2 (1-f)N/n]^{-1}$.  This increased resistance to stretching will lead to a greater polyhedral distortion of the AB interface.  This interfacial distortion is apparent in the oblate shape of the A domains of A15 phase in Figure \ref{fig: figure1} (b), corresponding to the oblate shape of Voronoi cells of lattice sites on the faces of the $Pm\bar{3}n$ unit cube.  For larger values of $n$ interfaces approach the shape of Voronoi cell at smaller values of $f$ and thus the BCC-A15 transition occurs at lower $f$.  Of course, for the inverted micelles we expect the opposite to be true; the tension of the highly stretched inner domains will prefer a spherical interface, ignoring the shape of the Voronoi cell.  Therefore, we would not expect that the inverse A15 phase is stable at large $f$ and
we do not find it within SCFT.

At low segregations, near the order-disorder transition (ODT), the predicted phase behavior is dramatically altered from the case of linear diblocks.  Generally, the ODT is shifted up to higher $\chi N$, or lower temperature.  For symmetric diblocks the critical point, indicating a mean field, second-order disordered (Dis) to lamellar transition, occurs at $\chi N = 10.495$ and $f=0.5$ \cite{mat_sch_prl}.  The critical points of the miktoarm star copolymer melts are shifted, for example,  $\chi N = 13.47$ and $f=0.559$ for $n=2$ and $\chi N = 13.94$ and $f=0.588$ for $n=3$.  The upward $\chi N$ shift in the ODT is consistent with other SCFT calculations for conformationally asymmetric copolymers \cite{vavasour_macro_93, mat_sch_macro_94_1, mat_bates_jpsci} and indicates that chain fluctuations near the ODT are systematically altered by molecular asymmetry.
Finally, we note that neither the cubic double-diamond nor the hexagonal-perforated lamellar phases are stable in these systems as with symmetric, linear diblocks.

\subsection{Strong Segregation}

In order to compute the strong segregation phase behavior of AB$_n$ miktoarm star copolymer melts, we compute the Dis-BCC, BCC-Hex, and Hex-Lam phase boundaries at $\chi N \approx 100$ \cite{boundary_note}.  Due to the numerical difficulties of considering the Gyr and A15 phases for large $\chi N$ we do not include these phases in our stability analysis.  Additionally, we ignore the window of stable closed-packed spheres, FCC, which is predicted by SCFT calculations to occur near the ODT.  Since we expect that the free-energy differences between the sphere phases will be relatively small at these strong segregations \cite{olm_mil_macro_98, gra_didon_kam_03} our calculation should capture the change from spheres to cylinders.  Likhtman and Semenov developed a general SST calculation to assess the stability of bicontinuous phases in the limit $\chi N \rightarrow \infty$.  Their calculation shows that the Gyr phase is unstable when chain fluctuations are ignored \cite{likhtman_macro_97}.  It is not known whether the bicontinuous Gyr phase is stable for finite but large $\chi N$ even within the well studied SCFT phase diagram of symmetric, linear diblocks \cite{mat_bates_macro_96_2}.  Recent experiments on melts of fluorinated PI-PEE diblocks indeed suggest that this morphology is an equilibrium phase for $\chi N \sim 100$ \cite{dav_hill_lodge_03}, and we expect that the same may be true for more asymmetric copolymers, such as these miktoarm stars.  Although we cannot prove its stability with our SCFT calculations, we expect that the Gyr phase appears at compositions intermediate to the stable Hex and Lam regions.

\begin{figure}\center
\epsfig{file=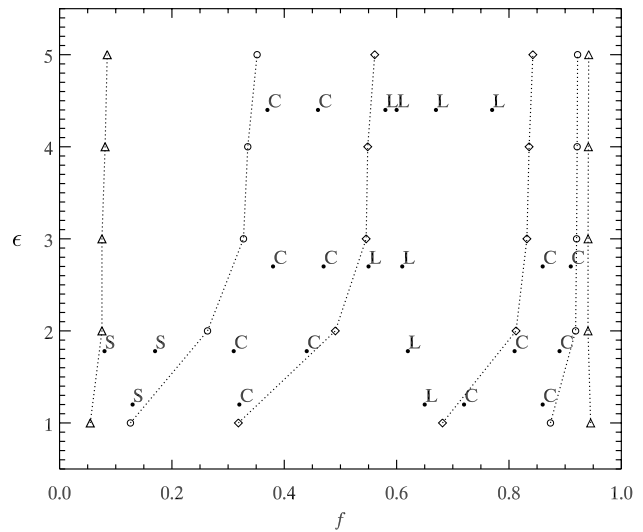,width=3.25in}\center \caption{SCFT results for AB$_n$ miktoarm stars at  large $\chi N$.  For our calculation, the asymmetry parameter $\epsilon=n_B=n$, the number of B blocks per molecule.  Triangles, $\triangle$, depict Dis-BCC transitions; open circles, $\circ$, depict BCC-Hex transitions; and diamonds, $\diamond$, indicate Hex-Lam transitions.  All boundaries are computed at $\chi N = 100$ with the exception of the low-$f$ BCC-Hex and Hex-Lam boundaries for $n=3$, 4 and 5. For $n=3$ these boundaries are computed at $\chi N =80$, and for $n=4$ and 5 these boundaries are computed at $\chi N =60$.  Dark circles, $\bullet$, indicate equilibrium results from experiments on PI-PS melts \cite{yang_macro_01}.  S, C, and L label spherical, cylindrical, and lamellar morphologies, respectively.} \label{fig: figure4}
\end{figure}

In Figure \ref{fig: figure4} we compare our SCFT results for AB$_m$ copolymers to the experimental results on polyisoprene (PI)-polystyrene (PS) A$_2$B$_2$ \cite{beyer_jpsci_99},
AB$_2$ \cite{pochan_macro_96}, AB$_3$
\cite{tselikas_jchemphys_96}, and AB$_5$ \cite{beyer_macro_99,
yang_macro_01}  miktoarm star copolymer melts.  Milner showed that SST phase boundaries depend only on $f$, and the asymmetry parameter, $\epsilon = (n_B/n_A) (a_A/a_B)$, where $n_B$ and $n_A$ are the number of A and B chains per molecule \cite{milner_macro_94}. Since we are considering symmetric monomers, $a_A=a_B$, $n_A=1$, and $n_B = n$ for our AB$_n$ copolymers, we have simply $\epsilon=n$.

We see from Figure \ref{fig: figure4} that SCFT very accurately captures the observed phase behavior of miktoarm star copolymer melts.  Therefore, ignoring composition fluctuations, as SCFT dictates, does not alter the phase behavior at this level of segregation.  Moreover, as the data suggest, the effect of the molecular asymmetry seems to saturate for $n>3$, and phase boundaries do not change significantly as a function of $f$ for further asymmetry $n$.  It has been suggested that the saturation of phase boundaries occurs when the spheres or cylinders
of the inner block reach close packing, {\sl i.e.} $f\approx 0.90$ for Hex and $f\approx 0.68$ for the body-centered cubic (BCC) lattice
\cite{yang_macro_01}.   In the next section we employ SST to explore how the symmetry of the micelle lattice frustrates their self-assembly, and moreover, how this frustration is related to the saturation of phase boundaries for $\epsilon \gg 1$.

\section{Strong-Segregation Theory:  ``Kinked-Path" Revisited}

The appearance of a stable A15 phase suggests that the shape of the AB interface is strongly affected by the lattice symmetry for highly asymmetric diblock melts.  Tension in the outer block chains maintains a uniform outer domain thickness and consequently distorts the interface into the shape of the Voronoi cell.  However, previous studies have demonstrated that for symmetric AB diblocks the shape of the interface, and therefore, the calculation of the free-energy are insensitive to the shape of the Voronoi cell \cite{mat_bates_macro_96, fredrickson_macro_93}.  The free energy is dominated by the tension of the AB interface, and so the minimal-area cylinder is favored.  Because of this, a 
unit-cell approximation (UCA) is often taken for the domain shape of the micelles.  For instance, in the columnar and spherical phases, the Voronoi cell is approximated by a perfect cylinder or sphere, respectively.  However, packing these cylinders or spheres into a space-filling lattice leaves voids in the interstices which is incompatible with  the incompressible melt state.  Therefore, the UCA can only provide an estimate for the free-energy of these morphologies.  In fact, this estimate is a lower-bound to the true free-energy since distorting the round, approximate unit cells can only raise the free energy either by stretching the outer block or distorting
the interface.  As our results in the last section suggest, the ``packing frustration'' \cite{mat_bates_macro_96} between the surface tension and the
stretching is highly dependent on molecular asymmetry.  It is also unlikely that the UCA can capture the saturation as a function
of the number of blocks, $n$; the close-packing limitation cannot be captured in the UCA since the inner volumes can fill 100\% of the approximate unit-cell without overlap \cite{yang_macro_01}.  

\begin{figure}\center
\epsfig{file=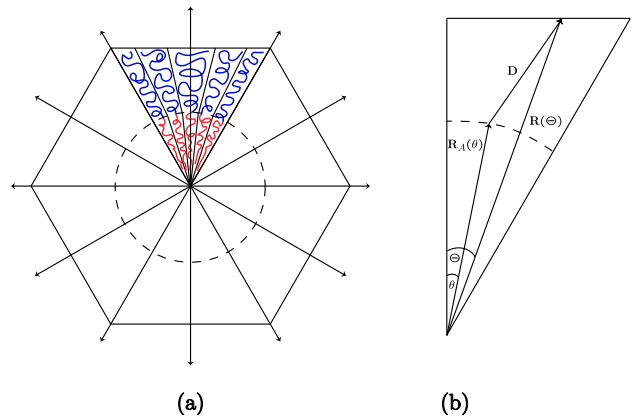,width=3.25in}\center \caption{In (a), a schematic representation of the kinked-path {\it ansatz} for a hexagonal unit cell with a circular interface dividing A and B domains, depicted with a dashed line.  Chains in the domain extend radially from the micelle center, while chains in the B region stretch away from the wall to the corners of the cell, modifying the shape of the outer domain wedges.  Symmetry dictates that  the B blocks extend radially along the six mirror planes of the unit cell, shown as arrows.   We need  only consider the bending of chain paths within the right-triangular wedge shown in (b).  Both angles, $\theta$ and $\Theta$, are measured with respect to the y-axis.  The A blocks extend along ${\bf R}_A (\theta)$, while the B blocks extend along ${\bf D}$, and ${\bf R} (\Theta)$ is the vector from the center of the micelle to the edge of the outer domain wedge at the wall of the unit cell.} \label{fig: figure5}
\end{figure}

In order to quantitatively explore the role of packing frustration in miktoarm star copolymer melts, we build upon the SST calculation of the Hex phase free energy for miktoarm star copolymers \cite{olm_mil_macro_98}.  Olmsted and Milner developed a ``kinked-path" {\it ansatz} for the extension of the copolymer chains, which allows the configuration to maintain a cylindrical AB interface while satisfying the constraints of melt incompressibility (see Figure \ref{fig: figure5}).  Here, we extend that calculation allowing the interface to adopt a more general class of interfaces, allowing us to systematically explore the effect of packing frustration in this morphology.

Within SST we ignore chain fluctuations, in addition to the composition fluctuations absent from the full SCFT.  We assume that each chain extends only along its classical trajectory, and these paths must be consistent with the constant monomer concentration of the melt state.  Following \cite{olm_mil_macro_98} we divide the hexagonal unit cell into infinitesimal wedges which extend along the direction of the chain paths.  Due to incompressibility we must have wedges with a volume fraction of A domain $f$ and a volume fraction of B domain $(1-f)$.  Unless the AB interface adopts the same shape as the Voronoi cell, the chains, and consequently the wedges must bend towards the cell corners in order to distribute the volume of the chains evenly.

Each chain starts with an A block that extends from the center of the micelle radially to the AB interface at an angle, $\theta$ (see Figure \ref{fig: figure5} (b)).  This path extends along the vector, $\mathbf{R}_A (\theta)$, which parameterizes the interface as function of $\theta$.  From that point on the interface, the outer blocks extend to some other point, at angle $\Theta$, on outer wall of the unit cell, along the vector $\mathbf{D} (\theta)$.  The vector which extends from the center of the micelle to the outer wall of the unit cell at $\Theta$ is given by $\textbf{R} (\Theta)$.   We can parameterize the kinking of the wedges by the function $\Theta(\theta)$ which maps the chain position at the interface to the position at the wall of the cell.  Therefore, the extension of the B portion of the wedge is given by $\textbf{D} (\theta)=\textbf{R}\big(\Theta(\theta)\big)-\textbf{R}_A (\theta)$.

\begin{figure}\center
\epsfig{file=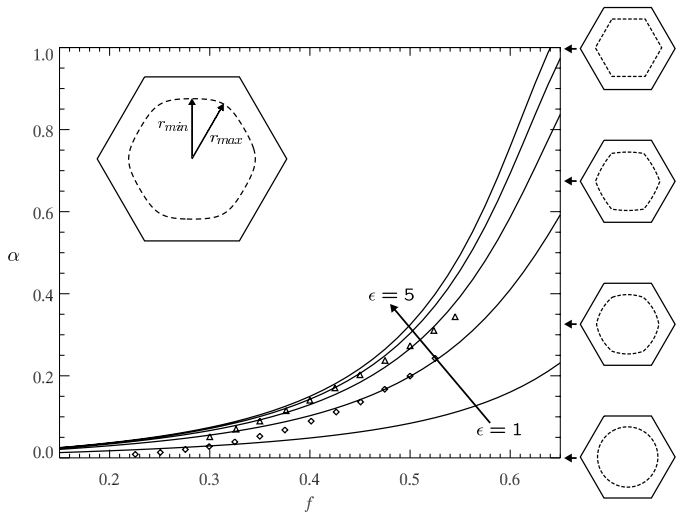,width=3.5in}\center \caption{Plots of the optimal interfacial shape parameter, $\alpha$, versus composition, $f$, for $\epsilon=1$, 2, 3, 4, and 5.  On the right are the interface shapes given by Eq. (\ref{eq: R_alpha}) for $\alpha=0, 1/3, 2/3,$ and 1, shown as dashed contours.  Inset on the upper left is the hexagonal unit-cell showing the AB interface (dashed line) extracted from the SCFT calculation for miktoarm stars with $n=3$, $\chi N =80$ and $f=0.54$. The minimum and maximum distances from the center of the micelle to the interface, $r_{min}$ and $r_{max}$ respectively, are labeled.  These are used to compute an estimate of the shape parameter for the SCFT calculations, $\alpha'$, defined in Eq.(\ref{eq: alpha'}).  These values of $\alpha'$ are depicted as diamonds, $\diamond$, for $n=2$ and $\chi N =100$, and triangles, $\triangle$, for $n=3$ and $\chi N =80$. } \label{fig: figure6}
\end{figure}

The hexgonal unit cell is parameterized for $0 \leq \theta \leq \pi / 6$ by
\begin{equation}
{\bf R} (\Theta) =  R_0 \sqrt{\frac{\pi}{2 \sqrt{3}}} \sec\Theta\,\hat{\bf r}(\Theta)\  ,
\label{eq: R}
\end{equation}
where $\pi R_0^2$ is the cross-sectional area of the micelle and $\hat{\bf r}(\beta)=\sin \beta \hat{\bf x}+ \cos \beta \hat{\bf y}$ is the radial
unit vector.  Because the configuration is strongly segregated and the inner domain is composed only of A monomers, we consider a class of interfaces which encloses a constant area fraction, $f$, of the total micelle area.  In particular, we would like to consider a class of interfaces which vary from area-minimizing and uniformly curved (circle) to stretch-minimizing and polygonal (hexagon).  Such a one parameter family of interfaces is: 
\begin{equation}
\label{eq:  R_alpha} {\bf R}_A (\theta; \alpha) = f^{1/2} R_0
\sqrt{(1-\alpha) + \alpha \Big(\frac{ \pi}{2 \sqrt{3}}\Big) \sec^2
\theta} \  \hat{\bf r}(\theta) 
\end{equation}
When $\alpha=0$ this is the circle and when $\alpha=1$ it traces out an affinely shrunken version of the hexagonal unit cell (\ref{eq: R}).  For all other values of $\alpha\in[0,1]$, the shape interpolates the two extremes (Figure \ref{fig: figure6}).  By construction the enclosed area of the interface is $f\pi R_0^2$ for all $\alpha$. 

We compute the free energy as a function of $\alpha$ and minimize in search of a variational ground state. We
expect the ground state to favor the minimal area, round interface ($\alpha=0$)
in the low $f$ limit because differences in stretching to the walls and corners of the Voronoi cell are small on the scale of the radius of gyration of the outer blocks, $ a \big[(1-f) N/n \big]^{1/2} $.  In the other limit, $f \rightarrow 1$, differences in chains stretching towards the corners and walls becomes large on this length scale, and the hexagonal interface ($\alpha=1$) will be preferred.  We refer to the former case as the round-interface limit (RIL) and the latter as the polyhedral-interface limit (PIL).  Of course, there is no
guarantee that the true ground state belongs to this class of
configurations.  Nonetheless, minimizing the free energy over
$\alpha$ will provide a variational upper-bound on the ground state free energy which is
lower than either that from the straight-path calculation or the circular interface calculation.

The essence of the kinked-path approximation is that each wedge must locally satisfy the volume constraint for the diblock so that:
\begin{multline}
f\left[{\bf R}(\Theta)-{\bf R}_A(\theta)\right]\times \frac{d}{d \theta} \left[{\bf R}(\Theta)+{\bf R}_A(\theta)\right]
= \\ (1-f) {\bf R}_A(\theta)\times \frac{d}{d \theta} {\bf R}_A(\theta)
\end{multline}
Given the parameterizations introduced in Eqs. (\ref{eq: R}) and (\ref{eq: R_alpha}) 
we find that
\begin{multline}
\label{eq: tantheta}
\tan \big( \Theta(\theta) \big) = \\ \frac{ (1-\alpha) \theta +
\alpha \big(\frac{ \pi}{2 \sqrt{3}}\big) \tan \theta -
\sqrt{\frac{\pi}{2 \sqrt{3}}} \Big( \frac{R_A (\alpha,\theta)}{R_0} \Big) \sin \theta }
{\big(\frac{ \pi}{2 \sqrt{3}}\big)-
\sqrt{\frac{\pi}{2 \sqrt{3}}} \Big( \frac{R_A (\alpha,\theta)}{R_0} \Big) \cos\theta } \ ,
\end{multline}
where $R_A (\alpha, \theta)=|{\bf R}_A (\alpha, \theta)|$.  It is easy to verify that $\tan \big( \Theta(\theta) \big)$ has the appropriate limits:  when $\alpha=0$, Eq. (\ref{eq: tantheta}) reduces to the results of ref. \cite{olm_mil_macro_98}, and when $\alpha = 1$, $\tan \big(\Theta(\theta)\big)=\tan \theta$, which is the straight-path {\it ansatz}.  

The free energy has three parts, arising from the interfacial energy, the stretching of the A blocks and the stretching of the B blocks:
\begin{equation}
F=F_{int}+F_{A}+F_{B}
\end{equation}
The interfacial energy per molecule (in units of $k_B T$) is given by the effective surface tension, $\gamma \sim \chi^{1/2}$ \cite{helfand_sapse_75}, times the area of the AB interface divided by the number of chains per micelle:
\begin{equation}
\label{eq: F_int}
F_{int}=\frac{2 N f^{1/2}}{\rho_0} \frac{\gamma \mathcal{A} }{ R_0} \ ,
\end{equation}
where $\pi R_0^2$ is the cross sectional area of the micelle and,
\begin{eqnarray}
\label{eq: A(alpha)}
\nonumber
\mathcal{A}(\alpha) &\!\!=\!\! & \frac{12
\int_0^{\pi/6} \! d\theta \big|\frac{d {\bf R}_A}{d
\theta}\big|}{2\pi R_0 f^{1/2}}  \\ &\!\! =\!\! & \frac{6}{\pi R_0 f^{1/2}}
\int_0^{\frac{\pi}{6 }} \!\! d\theta \sqrt{ R_A^2
+\bigg|\frac{dR_A}{d\theta}\bigg|^2} \ ,
\end{eqnarray}
is the area of the AB interface measured in units of the area of circular interface enclosing the same volume.  We compute the stretching free-energy using the self-consistent parabolic brush potential for melts \cite{mil_wit_cat_euro, mil_wit_cat_macro} as is shown in ref. \cite{olm_mil_macro_98}.  For the inner A domain we have
\begin{equation}
F_{A}= \frac{\pi^2 n_A^2}{16 N a_A^2} \mathcal{I}(\alpha) R_0^2 \  ,
\end{equation}
where 
\begin{equation}
\label{eq: I(alpha)}
\mathcal{I}(\alpha)=\frac{12
\int_0^{\pi/6} \! d\theta R_A^4 (\theta,\alpha)}{2\pi R_0^4 f^2}= (1-\alpha^2)
+ \frac{5 \pi}{9 \sqrt{3}}\alpha^2 \  ,
\end{equation}
is the reduced stretching moment of the A domain, introduced in ref. \cite{gra_didon_kam_03}.  The stretching term due to the outer B domain is given by
\begin{equation}
F_{B}=\frac{3 \pi^2 n_B^2}{8(1-f)^2N a_B^2} \mathcal{S}(\alpha,f) R_0^2 \ ,
\end{equation}
where the stretching moment of the B blocks is given by the integral
\begin{multline}
\label{eq: S(alpha,f)}
\mathcal{S}(\alpha,f) = \\  \int_0^{\frac{\pi}{6}} \!\! d \theta \ {\bf D}(\theta)
\cdot \bigg(\hat{{\bf N}}_A(\theta) \bigg|\frac{d {\bf R}_A}{d
\theta}\bigg| + 3\hat{{\bf N}}\big( \Theta(\theta)\big) \bigg|\frac{d {\bf R}}{d
\theta}\bigg|\bigg) \frac{D^2(\theta)}{\pi R_0^4} \ ,
\end{multline}
with
\begin{equation}
\label{eq: NdR}
\hat{{\bf N}}
\big(\Theta(\theta)\big) \bigg|\frac{d {\bf R}}{d \theta} \bigg| = R_0
\sqrt{\frac{\pi}{2\sqrt{3}}} \bigg|\frac{d \tan \Theta}{d
\theta} \bigg| \hat{{\bf y}} \ ,
\end{equation}
and
\begin{equation}
 \label{eq: NAdRA}
 \hat{{\bf N}}_A (\theta) \bigg|\frac{d {\bf R}_A}{d
\theta}\bigg| = R_A(\theta) \hat{{\bf r}} -
\frac{d R_A}{d \theta} \hat{\bf \theta} \ ,
\end{equation}
with $\hat{\theta}=\cos \theta \hat{\bf x} - \sin \theta \hat{\bf y}$. 

Combining these terms we have the total free-energy per molecule and minimizing over the dimension of the micelle, $R_0$, we compute the micelle free-energy per chain in units of $k_B T$:
\begin{equation}
\label{eq: F(alpha)} F_{Hex}(\alpha) = F_0 
\mathcal{A}(\alpha)^{2/3}
 \bigg[\frac{2 f \mathcal{I}(\alpha)}{ \epsilon} + \frac{12 \epsilon f \mathcal{S}(\alpha,f)}{(1-f)^2}
\bigg]^{1/3} \ ,
\end{equation}
where $\epsilon = (n_B/n_A)(a_A/a_B)$ and $F_0$ is proportional to $( \chi N )^{1/3}$and is independent of composition \cite{olm_mil_macro_98}.  

The inner domain stretching, $\mathcal{I} (\alpha)$, and the interfacial area, $\mathcal{A} (\alpha)$, are minimized by circular interface at $\alpha=0$, while the outer domain stretching, $\mathcal{S}(\alpha,f)$, is always minimized for $\alpha\geq 1$ (though we are only considering
$\alpha\in[0,1]$).  As $f \rightarrow 1$ the only finite energy configuration is the  hexagonal interface with $\alpha =1$.  Minimizing over alpha for $0.15>f>0.65$ and $1>\epsilon>5$ we find the optimum energy interface configurations and show the results in Figure \ref{fig: figure6}.  In the low $f$ limit a round interface is favored due to the relatively lower tension in the B chains.  For larger $f$, the stretching term of the B chains (\ref{eq: F(alpha)}) begins to dominate, distorting the interface towards the hexagon.  Because of the $O(\epsilon^2)$ increase in the effective spring of the outer chains, it follows that for larger asymmetry the onset of this polyhedral distortion occurs at lower compositions $f$.  

For high molecular asymmetries, $\epsilon \geq 3$, and A volume fractions larger than $f  \approx 0.65$, we find equilibrium shapes with  $\alpha > 1$.  In this case, the interface is bowed in, away from the walls, and pulled out towards the corners.  However, because it is the only finite energy configuration, each solution approaches $\alpha =1$ as $f \rightarrow 1$.  We expect that the true ground state prefers an interface which asymptotically approaches a hexagonal shape, rather than ``overshooting" $\alpha=1$ before returning to the $f \rightarrow 1$ limit.  The bowing inward of the interface is due to the oversimplified representation of the interface which does not allow the corners of the interface to relax smoothly towards the corners of the Voronoi cell.  Nevertheless, this simple representation of the interface shape proves sufficient to capture the effect of the packing frustration introduced by the hexagonal symmetry of the micelle lattice.

In order to make contact with the results of SCFT and experiment, we need to estimate
the value of $\alpha$ for an arbitrary interface.  We introduce a numerical measure of the distortion from the circular shape.  For copolymer melts in the large $\chi N$ limit the ratio of the interfacial thickness to the domain size scales as $(\chi N)^{-2/3}$ \cite{matsen_jphys_02}; thus, in the strong segregation limit the interface becomes infinitely thin on the scale of the domain.  We find the location of the interface by the contour for which the volume fraction of A is 0.5.  One of these contours is shown in the inset of Figure \ref{fig: figure6}.  Using this contour we measure the minimum and maximum distance from the center of the micelle to the interface, $r_{min}$ and $r_{max}$, respectively.  As a measure of the distortion from a circular interface, we
introduce
\begin{equation}
\label{eq: delta}
\delta  \equiv  \frac{r_{max}-r_{min}}{r_{max}+r_{min}} \ .
\end{equation}
which is similar to the normalized amplitude of the $\cos(6 \theta)$ modulation of the interface  \cite{mat_bates_macro_96}.  Since $\delta=0$ for the circular interface and $\delta=7 - 4 \sqrt{3} \approx 0.0718$ for a hexagonal interface, we may define
\begin{equation}
\label{eq: alpha'}
\alpha'  \equiv \frac{\delta}{7-4\sqrt{3}} \ .
\end{equation}
In the two limits $\alpha=0=\alpha'$ and $\alpha=1=\alpha'$ and 
it can be shown that $(\alpha'-\alpha)/\alpha$ is never greater than $5\%$ for the class of interfaces parameterized by $R_A (\alpha, \theta)$.  In particular, $(\alpha'-\alpha)/\alpha \approx 0.05, 0.04, 0.03, 0.02, 0.01$, and $0$ for $\alpha=0,0.2,0.4,0.6,0.8$, and $1$, respectively.

By using SCFT for linear diblock copolymers, Matsen and Bates showed that $\alpha' \approx 0.006$ for $f$ as large as $0.34$ \cite{mat_bates_macro_96}, and they concluded that
there is negligible distortion so that the effect of packing frustration can be ignored for the Hex phase.  However, we find that even for the $n=1$ diblock case, interfacial distortion becomes appreciable for $f \gtrsim 0.5$.  While the Hex phase is unstable at these compositions for AB diblocks, experiments for melts of ABC triblocks yield stable Hex phases with a composite AB inner domain volume fraction of about $0.65$ \cite{gido_macro_93}.  For these triblocks, the interface separating the outermost C domain and the B domain is found to be very nearly hexagonal, demonstrating that packing-frustration in the Hex phase is indeed amplified as the volume fraction of the inner domain is increased. Indeed, this frustration can be relieved by adding C-type homopolymer which aggregates in the corners of the hexagonal unit cell, allowing the BC interface to relax to uniform curvature \cite{lescanec_macro_98}. 

\begin{figure}\center
\epsfig{file=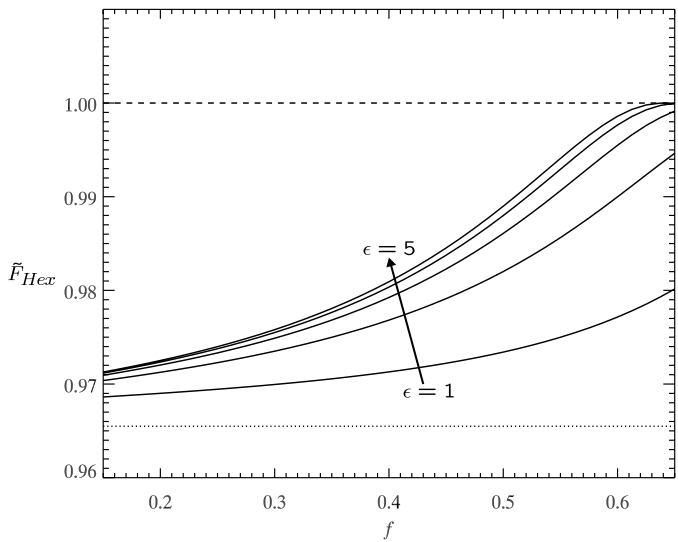,width=3.5in}\center \caption{Plots of the free-energy of the optimal configurations in units of the free-energy of the PIL, straight-path configuration, $\tilde{F}_{Hex}=F_{Hex} (\alpha_{min}) / F_{Hex} (\alpha=1)$.  The dashed line shows the straight-path upper bound, where the Voronoi cell and interface are hexagonal.  The dotted line shows the UCA lower bound, where the unit-cell and interface are circular.} \label{fig: figure7}
\end{figure}

We plot the free-energy of the optimal configuration as a function of composition for the various asymmetries in Figure \ref{fig: figure7}.  For lower compositions the free-energy is nearer the UCA lower bound.  As the composition increases the free-energy approaches the PIL upper-bound.  Clearly, for large $\epsilon$ and large $f$ the free-energy of our configuration is best approximated by the straight-path upper bound \cite{olm_mil_macro_98}.  This suggests that for asymmetric copolymers, the AB interface of the Hex micelle is significantly deformed from the uniformly-curved configuration over compositions ranges where this phase competes for stability.  We review the consequences of this conclusion for the phase calculation in the next section.

\section{Self-Consistent Field Vs. Strong-Segregation}

We compare the results of our SST calculation for asymmetric miktoarm star copolymer melts to the ``more exact" results of our SCFT calculations for the Hex phase.  Since the SST approximation results from an asymptotic expansion of the full SCFT partition function in the $\chi N \rightarrow \infty$ limit \cite{goveas_macro_97}, we expect that SST and SCFT results should agree provided that we have used the approximately correct shape of the AB interface.  Strictly speaking we should consider our SST results to be an upper bound on the free-energy of the Hex phase since we did not minimize over all possible ground state interface and chain configurations.  Nevertheless, we argue that this bound is sufficiently close to the true ground state to capture the phase behavior of asymmetric miktoarm stars.

We use the correspondence between $\alpha$ and $\alpha'$ for $R_A (\alpha, \theta)$ as a basis for direct comparison of the predicted values from SST and the ``measured'' values of $\alpha'$ from SCFT.  A comparison is made in Figure \ref{fig: figure6} for
$n=2$ and $n=3$. Despite our simplified representation of interface shape, $R_A (\alpha, \theta)$, and the crude means of measuring the equilibrium SCFT shape parameter, the predicted $\alpha$ and computed $\alpha'$ compare remarkably well.  In particular, the SCFT results confirm that increasing $f$ and $\epsilon$ increases the equilibrium shape distortion.  There is a systematic tendency in SST to overestimate the hexagonal distortion for low inner domain volume fractions.  This can be attributed to the inadequacy of our interface parameterization, $R_A (\alpha, \theta)$, at low distortions where the AB interface is likely best represented as a constant radius with a superposed $\cos (6\theta)$ modulation \cite{mat_bates_macro_96}.  Nevertheless, the agreement at larger $f$ can be taken as evidence that we have accurately accounted for packing frustration for our SST Hex phase calculation.

\begin{figure}
\center
\epsfig{file=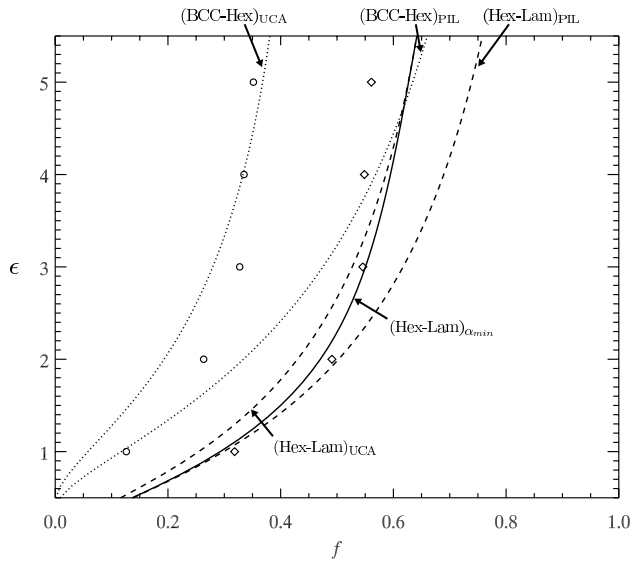,width=3.25in} \center \caption{The dashed line shows the predicted Lam-Hex phase boundaries computed using the PIL, upper-bound (leftmost boundary) and UCA, lower-bound (rightmost boundary) approximations for the SST Hex phase.  The solid line is the Lam-Hex boundary computed using the SST results of Section III.  The dotted lines are the Hex-BCC phase boundaries by comparing the PIL, upper-bound (leftmost) and UCA, lower-bound (rightmost) SST results for those morphologies.  For comparision the $\chi N \approx 100$ SCFT results for the Lam-Hex boundary are shown as diamonds, $\diamond$, and the BCC-Hex boundary are shown as open circles, $\circ$.} \label{fig: figure8}
\end{figure}

The effect of interfacial distortion on the location of the predicted Lam-Hex phase boundary for miktoarm star copolymers is shown in Figure \ref{fig: figure8}.  There we show (with dashed lines) the calculated phase boundaries using both the lower- and upper-bound approximations which assume that the interface and Voronoi cell are either both circular ({\sl i.e.} the UCA) or both hexagonal ({\sl i.e.} the PIL), respectively \cite{olm_mil_macro_98}.  Note that there is no difference between the round and polyhedral Voronoi cell for the Lam phase, and therefore, no approximation is necessary for this geometry.  Since the UCA calculation underestimates the Hex free-energy it overestimates the composition at which the Hex phase becomes unstable against
the Lam phase.  Likewise, the upper-bound calculation underestimates the location of the Lam-Hex phase boundary.  We expect, therefore, that the true Lam-Hex SST phase boundary should lie between these two approximate boundaries.

Though the upper-bound of the round-interface limit is close to the UCA lower-bound for symmetric diblocks \cite{olm_mil_macro_98}, the latter cannot be used to predict the phase behavior for asymmetric diblock melts in SST, as experimental results suggest \cite{yang_macro_01}.  This discrepancy is illustrated in Figure \ref{fig: figure8}:  while SCFT predicts that the effect of increasing  asymmetry saturates for $\epsilon \geq 3$, the UCA predicts that the phase boundaries move to larger $f$ as $\epsilon$ increases.  The solid line in Figure \ref{fig: figure8} shows the SST phase Lam-Hex phase boundary computed with the results of the previous section.  For $\epsilon \lesssim 2$ the AB interface is nearly undistorted and circular; thus, the free-energy is best approximated by the lower-bound, UCA calculation.  Hence, our Lam-Hex phase boundary follows the UCA phase boundary for low asymmetry.   However, for $\epsilon >  2$ the effect of interfacial distortion becomes important, and the free-energy is closely approximated by the upper-bound, PIL calculation.  Therefore, for large asymmetry our SST Lam-Hex boundary tends toward the hexagonal-interface approximation phase boundary.  This transition from the round to the hexagonal interface approximation is also reflected in the SCFT boundary.  Indeed, our SST and SCFT Lam-Hex phase boundaries agree well in this entire parameter range  ($1 \leq \epsilon \leq 5$).

Similarly, we see that Hex-BCC boundary computed from SCFT  tends to follow the UCA boundary for $\epsilon < 2$ and follows the polyhedral-interface approximation boundary for $\epsilon > 2$, confirming our picture of the effect of packing frustration on asymmetric copolymers.  Again, we know that interfacial distortion must be important along the phase boundary between sphere and columnar phases because the A15, with its minimal interfacial area, is the favored sphere phase there.  It would be interesting to pursue a one-parameter class of interfaces that interpolated between a spherical and polyhedral AB interface.  Such a calculation would not only predict the spherical-cylindrical micelle phase boundary but it would also elucidate the transition from BCC spheres with nearly spherical interfaces at small $f$ to A15 sphere-like micelles with polyhedral interfaces for larger $f$.

We can also see that within the polyhedral-interface approximation calculations, the effect of increasing asymmetry begins to saturate for $\epsilon > 5$, as seen in experiment \cite{yang_macro_01}.  Again, this stems from the relative $\epsilon^2$ increase in stretching energy of the B blocks over the A blocks.  In the large $\epsilon$ limit the A block stretching can be ignored, and (\ref{eq: F(alpha)}) can be approximated by $F_{Hex}(\alpha_{min}) \approx F_0 \big[ 12 \epsilon f \mathcal{A} (1)^2 S(1, f)/(1-f)^2 \big]^{1/3}$, where we have assumed that $\alpha_{min} \approx 1$.    The same is true for the free-energy of other morphologies, namely $F_{X} \approx F_0 \epsilon^{1/3} \Phi_{X} (f)$, where $X=$ Lam, A15, BCC, {\it etc.} and $\Phi_{X} (f)$ is some function specific to the morphology which depends only on composition \cite{olm_mil_macro_98}.  Because all free-energies scale the same way with $\epsilon$, the location of the transition between any two phases depends only on composition and is independent of $\epsilon$ in this limit.  
It should be noted that the effect of increasing asymmetry saturates for the UCA bounds as well,  though this occurs very close to $f=1$ for both the Lam-Hex and Hex-BCC boundaries.  The location of phase boundaries computed using the polyhedral-interface approximation saturates near $f \approx 0.50 $ and $f \approx 0.75$, for the Hex-BCC and Lam-Hex boundaries, respectively.

While the agreement between, experiment, our SCFT calculations and our SST analysis for $\epsilon \leq 5$ is improved over previous efforts, the SST results appears still to overestimate the effect of increasing asymmetry.  We should note, however, that SST is an asymptotic expansion of the full SCFT near $\chi N \rightarrow \infty$, and at finite values of $\chi N$ we need to assess the importance of higher order corrections.  It is known that lowest order, strong-segregation results are $O[(\chi N)^{1/3}]$ and that the leading corrections which can distinguish between phases are $O(1)$.  One of these corrections is associated with fluctuations of chain ends and junction points and are thus proportional to $(n_B + n_A -1)$ \cite{goveas_macro_97}.  The other is associated with a proximal layer near the interface, where the stretching energy of the chains deviates from the predictions of the parabolic brush potential \cite{likhtman_eurolett_00}.  At $\chi N \approx 100$ this correction can be as high as $20\%$, and will surely effect the relative free-energies of competing phases.  Further errors arise from the implicit
assumption of a parabolic chemical potential used to compute the free-energy associated with molten polymer brushes \cite{mil_wit_cat_euro, mil_wit_cat_macro}.  It has been shown that this approximation is equivalent to allowing for a negative chain end-density near the surface of a convex brush; nevertheless, it has been shown that relaxing the constraint of a non-negative end-distribution hardly perturbs the free-energy of a two-dimensional convex brush \cite{ball_marko_mil_wit}.  However small, this parabolic chemical-potential underestimates the constrained stretching free-energy of domains on the outside of cylindrical and spherical domains.  A correction to the stretching energy of B domains at large asymmetries of 1\% leads to a correction of the full free-energy of order 0.3\%.  While this seems small, the difference between upper- and lower-bound SST free-energies is only 3.6\%  and 6.8\%, for the Hex and BCC phases, respectively.  Since even small corrections to the free-energy lead to significant changes in the predicted phase behavior, we expect approximations of this order may be relevant.

\begin{figure}
\epsfig{file=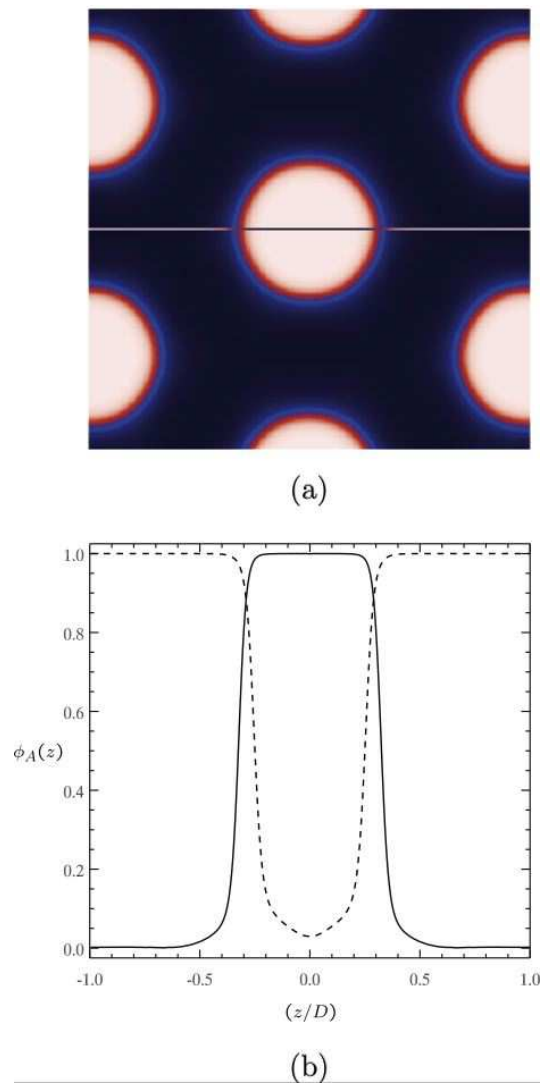,width=2.75in} \center \caption{Shown in (a), a real-space reconstruction of the A monomer volume fraction profile, $\phi_A (z)$, for the Hex phase at $n=4$, $\chi N =100$, and $f=0.3$.  High concentrations of A monomer are shown as bright red regions, and low concentrations of A monomer (high concentrations of B) are shown as dark blue regions.  In (b), the A monomer volume fraction profile along the horizontal line in (a) for $\chi N =100$ and $f=0.3$, the solid line, and $f=0.835$ (the inverse Hex phase), the dashed line. The equilibrium length scales, $D$, are $1.34 N^{1/2} a$ and $1.40 N^{1/2}a$, for $f=0.3$ and $f=0.835$, respectively.} \label{fig: figure9}
\end{figure}

\section{The Inverse Phases}
We have applied the SST analysis here to the Hex phase where the A domain composes the interior of the micelle and the multiple B blocks compose the outer domain.  These asymmetric configurations have the ``stiffer" blocks on the outside of micelle, hence, they are responsible for appreciable interfacial distortion.  However, we could also use our SST results to determine the phase behavior of the inverse micelles, where the multiple B blocks composed the inner domain.  Assuming that all B chains extend radially from the center of the micelle to the AB interface, we can compute the inverse micelle free-energy by the substitutions $\epsilon \rightarrow \epsilon^{-1}$ and $f \rightarrow (1-f)$ in Eq. (\ref{eq: F(alpha)}).  

We find, however, that such a calculation produces overestimates of the free-energy.  Moreover, the assumption that the B chains stretch only radially is inconsistent with SCFT results for the inverse phases.  Figure \ref{fig: figure9} (a) shows a real-space A monomer distribution for the Hex phase with $n=4$, $\chi N =100$ and $f=0.3$.   Figure \ref{fig: figure9} (b) plots the distribution of A monomer along a horizontal line extending across the micelle for $f=0.3$ and the inverse phase at $f=0.835$.  Clearly, both configurations are strongly segregated with the volume fraction of the A monomer, $\phi_A (z)$, constant outside the inner domain ($\phi_A (z)=0$ for $f=0.3$ and $\phi_A (z)=1$ for $f=0.835$).  However, while the volume-fraction of B component is zero inside the $f=0.3$ micelle, the volume fraction A component is never zero inside the inverse micelle at $f=0.835$.  This shows that in the inverse micelle, the miktoarm star junction points are not strictly confined to a narrow AB interface.  Rather, junction points are located somewhere within the inner domain and some of the chains must bend back towards the outer domain instead of stretching radially towards the center of the micelle.  
Thus the simple SST analysis of Section III does not apply to inverse phases since the assumption that junction points are confined to the AB interface is clearly violated.  In order to account for this we would  parameterize the distribution of junction points in the micelle, and minimize the free-energy over the distribution.  Moreover, we would need to find the optimal configuration of B blocks, bending inwards and outwards in the micelle.  Such a calculation introduces many more degrees of freedom; minimizing over these is likely very difficult.  Since SCFT provides numerically exact results for even very large segregations, it is a much better approach to finding the inverse micelle configurations to determining the phase behavior.

\section{Conclusion}

We have implemented SST and SCFT calculations which elucidate the coupling between the molecular asymmetry of diblock copolymers and packing frustration introduced by the micelle lattice.  In particular, we find that the shape of the AB interface is highly sensitive to molecular asymmetry, in the case of our miktoarm star copolymers, the number of B blocks per molecule.  For linear diblocks, or otherwise nearly symmetric architectures, we show that the AB interface of micelles in the Hex phase is rather insensitive to the shape of Voronoi cell and maintains a nearly constant mean curvature shape.  However, for AB$_n$ miktoarm star copolymers with $n\ge 3$ the interface is highly distorted towards the hexagonal shape of the Voronoi cell in regions where the Hex phase competes for stability.  The effect of this distortion is to shift the predicted Hex-Lam phase boundary to lower compositions than is predicted by approximating the Voronoi cell as a perfect cylinder.

As a consequence of the importance of interfaces, we expect the A15 phase of spherical micelles to be stable for highly asymmetric copolymers due to the minimal area of its Voronoi cell amongst  lattices in three-dimensions.  This prediction is borne out by our numerical SCFT calculations which compute the full phase behavior of AB$_n$ miktoarm star copolymers for $n \leq 5$.  Indeed, we find that as molecular asymmetry is increased, the stability of the A15 phase is enhanced indicating that the polyhedral approximation for the AB interface is more valid as the copolymer architecture becomes more asymmetric.

It is worth noting that the A15 phase of spherical micelles has yet to be identified experimentally in melts of highly asymmetric diblock copolymers.  Pochan, Gido and coworkers report the appearance of an equilibrium cubic phase of spheres in PS-PI AB$_2$ copolymer melts in regions of the $\epsilon-f$ phase space where we might expect A15 to be stable \cite{pochan_macro_96}.  However, they note that the small-angle X-ray scattering data from these melts cannot distinguish between a simple-cubic or BCC micelle lattice; nor does the data rule out the possibility of the A15 lattice which contains all of the reflections of the BCC lattice.  Experiments on sphere phases of more asymmetric miktoarm star copolymers in parameter ranges where A15 should be stable tend to yield poorly ordered arrangements of micelles \cite{gido_personal}.  Because the A15 and BCC phase are nearly degenerate (with a free-energy difference on the order of 0.2\%), the system may be kinetically trapped in some sort of glassy intermediate state in these parameter ranges.  Despite these difficulties, however, both the robustness of the minimal Voronoi cell area principle and the SCFT calculations strongly indicate that the A15 phase of spherical micelles is an equilibrium phase of copolymer melts of sufficient molecular asymmetry.

\section{Acknowledgments}
It is a pleasure to acknowledge stimulating discussions with B. DiDonna, G. Fredrickson, S. Gido, and V. Percec.  This work was supported by NSF Grants DMR01-29804, DMR01-02549, and INT99-10017, by the State of Pennsylvania under the Nanotechnology Institute, and by the University of Pennsylvania Research Foundation.


\begin{thebibliography}{99}
\bibitem{boundary_note}
All boundaries are computed for $\chi N =100$ with the exception of the low-$f$ BCC-Hex and Hex-Lam boundaries for $n=3$, 4 and 5.  For $n=3$ these boundaries are computed for $\chi N =80$, and for $n=4$ and 5 these boundaries are computed at $\chi N =60$.  As $\chi N \rightarrow \infty$ the phase boundaries become independent of the degree of segregation, $\chi N$, \cite{milner_macro_94} and thus, we expect that the phase boundaries computed at lower segregation should very closely approximate the higher segregation results.
\bibitem{FCC_note}
When considering close-packed lattices of spherical micelles, hexagonally close-packed and FCC lattices are known to have degenerate free-energies within mean field theory \cite{mat_bates_macro_96_2}.  Here, we consider only the free-energy of the phase of $Fm\bar{3}m$ symmetry, the FCC phase.
\end{thebibliography}
\end{document}